\documentclass[11pt]{article}

\usepackage{amsmath,amssymb,amsthm,amscd,latexsym}
\usepackage{hyperref}
\usepackage{cite}
\usepackage{graphicx,epsfig}
\RequirePackage{color}

\allowdisplaybreaks[1]

\def\ton#1{\left(#1\right)}
\def\qua#1{\left[#1\right]}
\def\grf#1{\left\{#1\right\}}

\def\nnb{\\ \nonumber \\}
\def\nb{\nonumber &}
\def\J2{J_2}

\def\ee{e^2}

\def\nk{n_{\rm b}}
\def\kap{\bds{\hat{k}}}
\def\kx{\hat{k}_x}
\def\ky{\hat{k}_y}
\def\kz{\hat{k}_z}

\def\rfr#1{eq. (\ref{#1})}

\def\dert#1#2{\frac{{{d}}{#1}}{{{d}}{#2}}}

\def\virg#1{``#1"}

\def\eqi{\begin{equation}}
\def\eqf{\end{equation}}
\def\eqia{\begin{eqnarray}}
\def\eqfa{\end{eqnarray}}
\def\Om{\mathit{\Omega}}
\def\rp#1#2{{#1\over#2}}
\def\lb#1{\label{#1}}
\def\bds#1{\boldsymbol{#1}}
\def\coo{\cos 2\omega}
\def\soo{\sin 2\omega}

\def\cI{\cos I}

\def\cII{\cos 2I}
\def\sII{\sin 2I}

\begin{document}

\title{A possible new test of general relativity with Juno}

\author{L. Iorio\\ Ministero dell'Istruzione, dell'Universit$\grave{\textrm{a}}$ e della Ricerca (M.I.U.R.) \\ Fellow of the Royal Astronomical Society (F.R.A.S.)\\ Viale Unit$\grave{\textrm{a}}$ di Italia 68, 70125, Bari (BA), Italy\\ email: lorenzo.iorio@libero.it}

\maketitle

\begin{abstract}
The expansion in multipoles $J_{\ell},\ell=2,\ldots$ of the gravitational potential of a rotating body affects the orbital motion of a test particle orbiting it with long-term perturbations both at a classical and at a relativistic level.  In this preliminary sensitivity analysis, we show that, for the first time, the $\J2 c^{-2}$ effects could be measured by the ongoing Juno mission in the gravitational field of Jupiter during its nearly yearlong science phase (10 November 2016-5 October 2017) thanks to its high eccentricity ($e=0.947$) and to the huge oblateness of Jupiter ($\J2=1.47\times 10^{-2}$). The semimajor axis $a$ and the perijove $\omega$ of Juno are expected to be shifted by $\Delta a \lesssim 700-900$ m and $\Delta\omega\lesssim 50-60$ milliarcseconds (mas), respectively, over $1-2$ year. A numerical analysis shows also that the expected $\J2  c^{-2}$ range-rate signal for Juno should be as large as $\approx 280$ microns per second ($\mu$m s$^{-1}$) during a typical 6 h pass at its closest approach. Independent analyses previously performed by other researchers about the measurability of the Lense-Thirring effect showed that the radio science apparatus of Juno should reach an accuracy in Doppler range-rate measurements of $\approx 1-5$ $\mu$m s$^{-1}$ over such  passes. The range-rate signature of the classical even zonal perturbations is different from the 1PN one. Thus, further investigations, based on covariance analyses of simulated Doppler data and dedicated parameters estimation, are worth of further consideration.
It turns out that the $\J2 c^{-2}$ effects cannot be  responsible of the flyby anomaly in the gravitational field of the Earth. A dedicated spacecraft in a 6678 km $\times$ 57103 km polar orbit would experience a geocentric $\J2 c^{-2}$ range-rate shift of $\approx 0.4$ mm s$^{-1}$.
\end{abstract}



\centerline

\section{Introduction}
The general theory of relativity  is one of the fundamental pillars of our knowledge of the natural world, being, to date, the best theory of the gravitational interaction at our disposal. Thus it is important to put it on the test in as much ways as possible to increase our confidence in it, especially in view of the extrapolations of its validity to extreme scenarios. In these cases, empirical checks are more difficult  and/or the interpretation of existing observations heavily rely upon more or less speculative assumptions concerning the history of the systems considered and the physics governing them in such  regimes.

In this respect, the weak-field and slow-motion approximation of general relativity has played so far an important role. If on the one hand, the magnitude of its predicted effects is generally modest with respect to those expected in the strong-field regime, on the other hand, they occur in the gravitational fields of stars and planets such as our Sun, the Earth and some of the major bodies of our Solar System which can be probed with great confidence with  man-made objects tracked with increasing accuracy. Moreover, the key physical parameters of such natural laboratories are, in general, well known from a variety of independent space-based missions, thus reducing the untested (and often untestable) speculations down to a minimum.

So far, the weak-field tests of general relativity probed the behaviour of the electromagnetic waves propagating in the vicinity of massive objects,  and the orbital dynamics of test particles and gyroscopes moving around them \cite{2001LRR.....4....4W}. As far as the latter class of effects are concerned, at the first Post-Newtonian (1PN) level they  are
\begin{itemize}
\item The precession of the pericenter of a test particle in the field of a spherically symmetric, static  source. It corresponds to the celebrated Einstein precession of the perihelion of Mercury \cite{1915SPAW...47..831E} of 42.98 arcseconds per century \cite{1986Natur.320...39N}.
\item The geodesic or de Sitter-Fokker precession of a gyroscope orbiting a spherically symmetric, static body \cite{desitter16,fokker21,2012EM&P..109...55R}. It was measured  in the field of the Sun by tracking the motion of the Earth-Moon system with the Lunar Laser Ranging (LLR) technique \cite{Mul97,Williams04,Mull08}, and in the field of the Earth with the man-made gyroscopes carried onboard the Gravity Probe B (GP-B) spacecraft \cite{Everittetal2011}.
\item The gravitomagnetic effects occurring in the vicinity of a stationary source: the Lense-Thirring orbital precessions of a test particle \cite{LT18} and the Pugh-Schiff precession of a gyroscope \cite{Pugh59,Schiff60b} orbiting a rotating body, both caused by the proper angular momentum $\bds S$ of the latter. A measurement of the Pugh-Schiff effect was recently made with the GP-B mission \cite{Everittetal2011}. For a recent overview of the attempts to measure the Lense-Thirring effect in the Earth's field, see, e.g., \cite{2013CEJPh..11..531R} and references therein.
\end{itemize}
One of the major sources of systematic bias in the attempts to detect such relativistic orbital effects is represented by the competing classical secular precessions induced by the expansion in multipoles \textcolor{black}{$J_{\ell}, \ell=2,4,\ldots$} of the Newtonian gravitational potential of the central body accounting for its departures from spherical symmetry. It turns out that the quadrupole of the gravitational field of the primary has an impact on the motion of a test particle \textcolor{black}{not only at a Newtonian level but} also at the 1PN level \cite{1963uasg.proc...69K,Krause64,1966MNRAS.131..483G,1978GReGr...9.1101M,1981PhRvD..24.2332B,1979lbge.book.....I}, inducing  non-vanishing long-term orbital perturbations {of order $\J2 c^{-2}$} on all the Keplerian orbital elements \cite{1988CeMec..42...81S,Soffel89,1990CeMDA..47..205H,1991ercm.book.....B,arabo}. So far, such effects did not receive much attention since it was always believed that they are too small to be detectable in any foreseeable future. For example, Brumberg \cite{1991ercm.book.....B} wrote that \virg{these effects are not observable as yet and [\ldots] can hardly have a great chance of being detected in the near future}. \textcolor{black}{Also the spin of a gyroscope orbiting an oblate body experiences a general relativistic $\J2 c^{-2}$ contribution to its precession \cite{1969NCimL...1..933O,1969Ap&SS...4..119O,1970PhRvD...2.1428B,1977PhRvD..16..946T,1988nznf.conf..685B}, in addition to the de Sitter-Fokker and to the Pugh-Schiff ones; in the present paper, we will not deal with it.}

The scope of  this paper is to show that recent developments in space sciences may soon overturn such opinions\textcolor{black}{, at least as far as the $J_2 c^{-2}$ orbital effects are concerned}. After reviewing the 1PN orbital effects due to the oblateness of the central body in Section \ref{prima}, the possibility of detecting them in some astronomical systems is considered in Section \ref{terz}. In Section \ref{juno} we will show that there are good perspective to accurately measure the $\J2 c^{-2}$ orbital effects in the gravitational field of Jupiter with the Juno spacecraft currently en route to it. Also the Earth scenario is examined in Section \ref{terra}. Section \ref{fine} summarizes our findings.
\section{The 1PN orbital effects of order $\J2 c^{-2}$}\lb{prima}
The direct Post-Newtonian perturbations due to the quadrupole moment $\J2$ of the central source of mass $M$ and equatorial radius $R$ were explicitly computed in alternative metric theories of gravity
by Soffel et al.\footnote{The rates of change of  \rfr{dadt1}-\rfr{dodt1} can be obtained by
taking the time derivatives of the long-periodic (LP) and secular (S) shifts $\Delta a,\Delta e,\ldots$ of the orbital elements calculated by Soffel
et al. \cite{1988CeMec..42...81S} and by using the general relativistic values of the parameters entering them.} \cite{1988CeMec..42...81S}
and by Brumberg\footnote{For a better comparison with the results by Brumberg, the following replacements
are to be made in his formulas at pag. 114-115 of \cite{1991ercm.book.....B}: $Q\rightarrow -\J2 GM R^2=-\J2\nk^2 a^3 R^2,
m\rightarrow GM/c^2, \alpha\rightarrow 0, \varpi\rightarrow\omega$. Here $c$ is the speed of light in vacuum, $G$ is the Newtonian constant of gravitation, and $\nk=\sqrt{GM a^{-3}}$ is the Keplerian mean motion of the test particle.} \cite{1991ercm.book.....B}
by means of the standard Gauss equations for the variation of the Keplerian orbital elements which are\footnote{In the following we will not consider the mean anomaly $\mathcal{M}$ since it is rarely considered in empirical studies. It is so because the Keplerian mean motion  should be subtracted from its rate; actually, the knowledge of $\nk$ is limited by the uncertainty in the primary's $GM$ and, in the case of artificial spacecrafts, by any non-gravitational perturbations plaguing the semimajor axis $a$.} the semimajor axis $a$, the eccentricity $e$, the inclination $I$ of the orbital plane to the reference $\{x,y\}$ plane, the longitude of the ascending node $\Om$ and the argument of pericenter $\omega$. In general relativity, they are, to first order,
\begin{align}
\dert a t \lb{dadt1} & = \rp{9\J2\nk^3 R^2\ee\ton{6+\ee}\sin^2 I\soo}{8 c^2\ton{1 - \ee}^4}, \nnb
\dert e t \lb{dedt1} & = \rp{21\J2\nk^3 R^2e\ton{2+\ee}\sin^2 I\soo}{16 c^2\ton{1 - \ee}^3}, \nnb
\dert I t \lb{dIdt1} & = \rp{3\J2\nk^3 R^2\ee\sII\soo}{8 c^2\ton{1 - \ee}^3}, \nnb
\dert \Om t \lb{dOdt1} & = \rp{3\J2\nk^3 R^2\cI\ton{6-\ee\coo}}{4 c^2\ton{1 - \ee}^3}, \nnb
\dert \omega t \lb{dodt1} \nb = -\rp{3\J2\nk^3 R^2}{32 c^2\ton{1 - \ee}^3}\grf{32-3\ee -2\ton{7+2\ee}\coo +\right.\nnb
&+\left. \cII\qua{48-9\ee + \ton{14-4\ee}\coo }},
\end{align}
We performed our own calculations within general relativity by using the Gauss equations and the true anomaly $f$ as fast variable of integration;
by using a Keplerian ellipse as unperturbed reference trajectory,  we were able to obtain \rfr{dadt1}-\rfr{dodt1}.

The indirect, second order mixed perturbations induced by the Newtonian quadrupole field and the Schwarzschild acceleration were computed in general relativity by Heimberger et al. \cite{1990CeMDA..47..205H}  by using the Lie transform perturbation theory \cite{1999thor.book.....B} with the Delaunay variables
$l=\mathcal{M},L=\nk a^2,
g=\omega,G^{'}=\nk a^2\sqrt{1-\ee},h=\Om,H=\nk a^2\sqrt{1-\ee}\cI$.
In terms of the usual Keplerian orbital elements, they are \cite{1990CeMDA..47..205H}
\begin{align}
\dert a t \lb{dadt2} & = 0, \nnb
\dert e t \lb{dedt2} & = \rp{9\J2\nk^3 R^2 e\sin^2 I\soo}{4c^2\ton{1 - \ee}^2}, \nnb
\dert I t \lb{dIdt2} & = 0, \nnb
\dert\Om t \lb{dOdt2} & = -\rp{3\J2\nk^3 R^2\cI\ton{7+29\ee +18\sqrt{1-\ee}-3\ee\coo  }}{4c^2\ton{1 - \ee}^3}, \nnb
\dert\omega t  \lb{dodt2}\nb = \rp{3\J2\nk^3 R^2}{16 c^2\ton{1 - \ee}^3}\grf{87 e^2+3 \left(2+e^2\right) \coo+\right.\nnb
\nb + \left.\cII\left[145 e^2-3 \left(2+5 e^2\right) \coo+108 \sqrt{1-e^2}+107\right]+\right.\nnb
& + \left.60 \sqrt{1-e^2}+45}.
\end{align}
For another computation based on the Lie transform perturbation theory and the Delaunay elements, see \cite{arabo}.

Both \rfr{dadt1}-\rfr{dodt1} and \rfr{dadt2}-\rfr{dodt2} hold in a frame whose reference $z$ axis is
aligned with the unit vector $\kap$ of the spin axis of the central body.
\section{Perspectives for a detection}\lb{terz}
In this Section we look for some astronomical scenarios which may allow for a measurement of the $\J2 c^{-2}$ orbital effects. Contrary to what one might think at first glance, such a perspective is not unrealistic.
\subsection{Jupiter and Juno}\lb{juno}
Juno\footnote{See http://www.nasa.gov/mission$\_$pages/juno/main/index.html on the WEB.} \cite{2007AcAau..61..932M} is a spacecraft en route to Jupiter where its arrival is scheduled for July 2016. Its science phase has a nominal duration of almost 1 yr (10 November  2016-5 October 2017) \cite{2011Icar..216..440H}. During it, Juno will move along a highly elliptical ($e=0.947$), polar ($I=90$ deg\footnote{In a coordinate system whose reference $z$ axis is aligned with the Jupiter's spin axis.}) orbit to accurately map, among the other things, the gravitational field of Jupiter  \cite{2004DPS....36.1409A} through the perturbations on the Juno's wide ($a=20.03 R$, $P_{\rm b}=11.07$ d) trajectory due the departures from spherical symmetry of the Jovian gravitational potential.

In view of the high diurnal rotation rate of Jupiter and of the peculiar orbital geometry of Juno, it was suggested \cite{2010NewA...15..554I} that it could also carry out a  measurement of the Lense-Thirring effect at  a percent level yielding a determination of the angular momentum \textcolor{black}{$\bds S$} of Jupiter; further independent analyses confirmed such a possibility \cite{2011AGUFM.P41B1620F,2011Icar..216..440H}. From a practical point of view, the measurement would be conducted with the Juno's radio science system\footnote{While the Ka-Band system is for gravity science, the X-Band apparatus is used for the spacecraft orbit determination and navigation \cite{2011Icar..216..440H}. As far as the Doppler range-rate measurements are concerned, the X-Band system is less accurate than the Ka-Band one by a factor of about $10-100$ \cite{2011Icar..216..440H}.} (X-Band, Ka-Band)  providing an accurate determination of the Doppler shift \cite{2011AGUFM.P41B1620F,2011Icar..216..440H} during many\footnote{It depends on the pointing of the Juno's high-gain antenna \cite{2011Icar..216..440H}; if it does not point to the Earth, as in the microwave radiometry passes when it will point to Jupiter, Doppler tracking is impossible. According to Helled et al. \cite{2011Icar..216..440H}, about 22 useful perijove passes  should be  available during the scheduled science phase.} of the scheduled 31 orbits in the science phase, mainly at the perijove passages lasting about 6 hr. Indeed, as pointed out in \cite{2011AGUFM.P41B1620F},  large longitude-keeping maneuvers would compromise the dynamical coherence of the orbit, making, thus, practically infeasible an analysis based on steady time series of the Keplerian orbital elements. The expected overall range-rate accuracy should be of the order of $1-5\ \mu$m s$^{-1}$ over time scales of $10^3$ s \cite{2010NewA...15..554I,2011AGUFM.P41B1620F,2011Icar..216..440H} for the Ka-Band apparatus, while the spacecraft's position and velocity should be known with an uncertainty of about 10 m and 1 mm s$^{-1}$, respectively, in the three spatial directions \cite{2011Icar..216..440H}.

The high Jupiter's oblateness ($\J2= 1.47\times 10^{-2}$), combined with the relevant eccentricity of the Juno's orbit and with its accurate radio science Ka-Band system, may offer a unique opportunity to detect, for the first time, the $\J2 c^{-2}$ orbital effects. Here we will offer just a preliminary sensitivity analysis; in view of its promising outcome, it is hoped that it will prompt further, more detailed investigations by independent teams in such a way it occurred for \cite{2011AGUFM.P41B1620F,2011Icar..216..440H}.

As a first outlook, in Figure \ref{plotti} we depict the  shifts $\Delta a$ and $\Delta\omega$ of the semimajor axis and of the perijove of Juno as  functions of the initial value $\omega_0$ of the perijove itself and of the duration $T$ of the science phase.  We obtained $\Delta a$ and $\Delta\omega$ by integrating \rfr{dadt1} and the sum of \rfr{dodt1} and \rfr{dodt2} from $t=0$ to $t=T$. We posed $\omega = \omega_0 + \dot\omega t$, where $\dot\omega$ is the sum of both the classical $\J2$ and of the Schwarzschild+Lense-Thirring perijove precessions. Although the nominal duration of the science phase of Juno is about 1 year, we considered also an extension of it, as it is not infrequent in successful space missions.
\begin{figure*}
\centering
\begin{tabular}{cc}
\epsfig{file=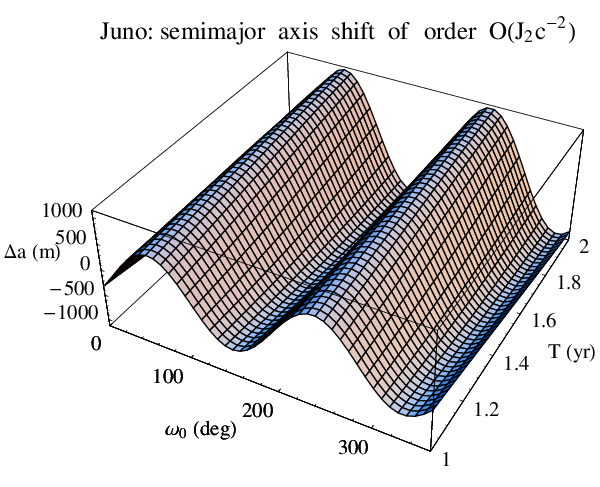,width=0.50\linewidth,clip=} & \epsfig{file=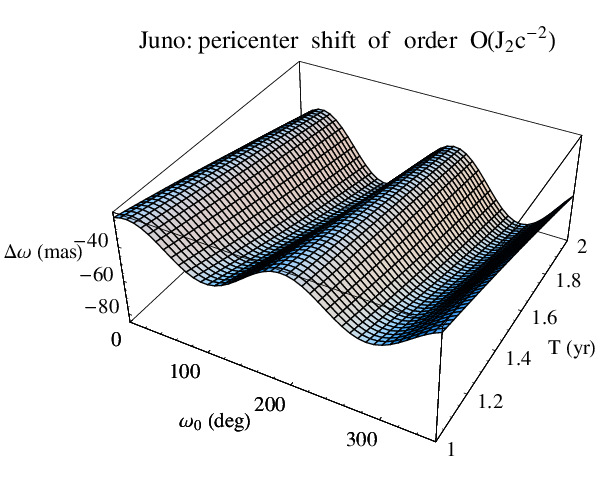,width=0.50\linewidth,clip=}\\
\epsfig{file=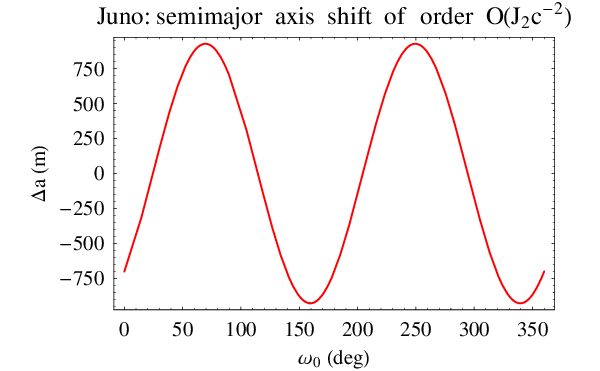,width=0.50\linewidth,clip=} & \epsfig{file=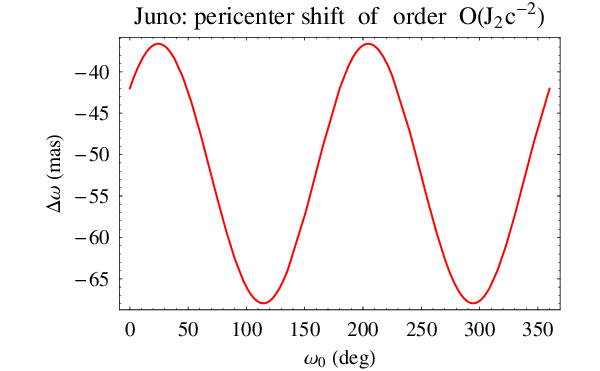,width=0.50\linewidth,clip=}\\
\epsfig{file=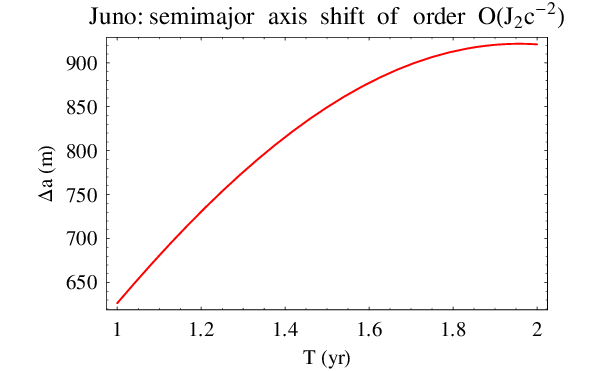,width=0.50\linewidth,clip=} & \epsfig{file=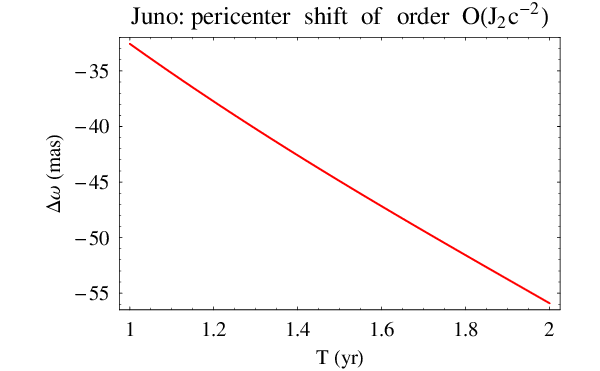,width=0.50\linewidth,clip=}\\
\end{tabular}
\caption{Left column, upper panel: plot of the semimajor axis shift $\Delta a$, in m, of Juno as a function of the initial value $\omega_0$ of the perijove and of the duration $T$, in yr, of the science phase of the mission. Left column, middle panel: plot of the semimajor axis shift $\Delta a$, in m, of Juno as a function of the initial value $\omega_0$ for $T=1.6$ yr. Left column, lower panel: plot of the semimajor axis shift $\Delta a$, in m, of Juno as a function of the duration $T$ of the science phase of the mission for $\omega_0=60$ deg. Right column, upper panel: plot of the perijove shift $\Delta \omega$ of Juno, in mas, as a function of the initial value $\omega_0$ of the perijove itself and of the duration $T$, in yr, of the science phase of the mission. Right column, middle panel: plot of the perijove shift $\Delta \omega$ of Juno, in mas, as a function of the initial value $\omega_0$ for $T=1.6$ yr. Right column, lower panel: plot of the perijove shift $\Delta \omega$ of Juno, in mas, as a function of the duration $T$ of the science phase of the mission for $\omega_0=60$ deg.}\lb{plotti}
\end{figure*}
The shift of the semimajor axis can reach values as large as $\Delta a\approx 600-900$ m, while the perijove shift can amount to $|\Delta\omega|\approx 50-60$ milliarcseconds (mas); from such a point of view, an extension of the science phase up to $\approx 1.5-2$ yr would be certainly desirable. We did not show the eccentricity  shift $\Delta e$ since it turns out to be as little as $\approx 5$ mas; the node and the inclination are not affected at $\J2 c^{-2}$ level because of the polar orbital configuration. It is important to note that, contrary to the perijove, the semimajor axis does  experience neither long-periodic nor secular competing perturbations of gravitational origin; thus, there would be no further gravitational aliasing bias on $a$. As it can be inferred from \rfr{dodt1} and \rfr{dodt2}, the perijove has both a secular and a long-periodic $\J2 c^{-2}$ component which would help in separating it from the biasing classical and relativistic (Schwarzschild+Lense-Thirring) secular precessions; according to \cite{2010NewA...15..554I}, the mismodelling in the Newtonian perijove precessions due to the even zonals of the Jupiter's gravitational potential would be as little as $\approx 1$ mas yr$^{-1}$.

In order to make a closer connection with the actual direct observable, in the right panel of Figure \ref{rrate} we plot a numerically integrated Earth-Juno range-rate signal $\Delta\dot\rho$ due to the $\J2 c^{-2}$ acceleration over a 6 hr pass centered about the first perijove passage after the beginning of the science phase.
\begin{figure*}
\centering
\begin{tabular}{cc}
\epsfig{file=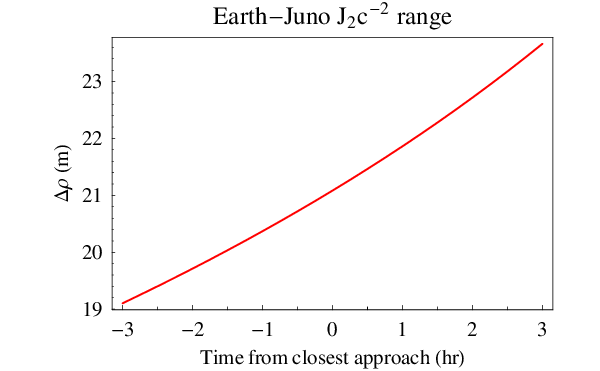,width=0.45\linewidth,clip=} & \epsfig{file=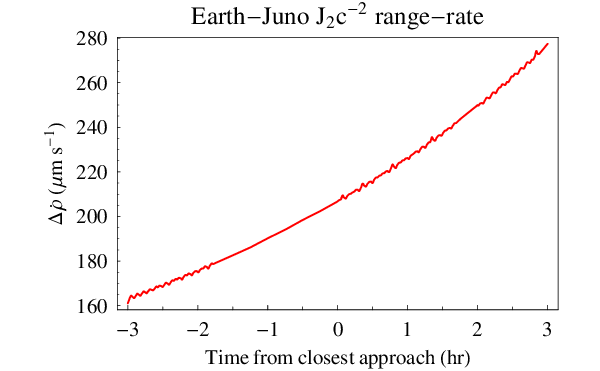,width=0.45\linewidth,clip=} \\
\end{tabular}
\caption{From the left to the right: Earth-Juno range, in m, and range-rate, in $\mu$m s$^{-1}$, due to the $\J2 c^{-2}$ acceleration. Both were calculated by numerically integrating the equations of motion of the Earth, Jupiter and Juno with and without the $\J2 c^{-2}$ term over 1 yr in a coordinate system with the ICRF equator as reference $\{x,y\}$ plane. In it, the Jupiter' spin axis is $\kx=-0.0146021,\ky=-0.430337,\kz=0.90255$, as it can be inferred from the ICRF equatorial coordinates at epoch J$2000.0$ of its north pole of rotation \cite{2007CeMDA..98..155S}.  Both the integrations shared the same initial conditions corresponding to the beginning of the science phase (10 November 2016) \cite{2011Icar..216..440H}; for Jupiter and the Earth they were retrieved from the WEB interface HORIZONS by NASA/JPL. For Juno we adopted $x_0=-0.0444405\ R,y_0=-1.3097\ R,z_0=2.74684\ R,\dot x_0=19.4951\ {\rm km\ s^{-1}}, \dot y_0=-11.9024\ {\rm km\ s^{-1}},\dot z_0=23.5485\ {\rm km\ s^{-1}}$ corresponding to an initial polar orbital configuration with $\omega^{'}_0=5.7$ deg with respect to the Jupiter's equator \cite{2011Icar..216..440H}. The time interval of the plots, covering 6 hr, is  centered about the first perijove passage after the beginning of the science phase.}\lb{rrate}
\end{figure*}
The mathematical model of the $\J2 c^{-2}$ acceleration used in our numerical integration of the equations of motion of Juno has no restrictions on the spatial orientation of $\kap$. This is a useful feature since it allows to use different coordinate systems. It is likely that the real data will be actually  processed in the ICRF frame; in it, the Jupiter's spin axis is $\kx=-0.0146021,\ky=-0.430337,\kz=0.90255$, as it can be retrieved from the celestial coordinates of its north pole of rotation at epoch J$2000.0$  \cite{2007CeMDA..98..155S}.
The resulting $\J2 c^{-2}$ range-rate signal amounts to about 280 $\mu$m s$^{-1}$, thus supporting the feasibility of our proposed measurement.
From the left panel of Figure \ref{rrate}, we also note  that the $\J2 c^{-2}$ range shift $\Delta\rho$ of Juno is as large as $\approx 20$ m over the same interval.

We stress once again that this is just a sensitivity analysis aimed to explore the possibility of a detection of the effect considered. It should be accompanied by a dedicated full covariance study with the simulated data of the real Doppler range-rate measurements at the perijove passages, and by the estimation of dedicated solve-for parameters explicitly accounting for the $\J2 c^{-2}$ acceleration in the dynamical models. Nonetheless, the encouraging outcome of the investigations in \cite{2011AGUFM.P41B1620F,2011Icar..216..440H} concerning the Lense-Thirring effect makes us confident that also for the $\J2 c^{-2}$  effect a percent determination with Juno may be possible.

\textcolor{black}{Here we give just a concise list of potentially competing orbital effects whose impact on the proposed measurement may be the object of future dedicated investigations. The odd zonals $J_\ell, \ell=3,5,\ldots$ should, in principle, be considered in more detail with respect to \cite{2010NewA...15..554I}. Indeed, if, on the one hand, they do not produce long-term perturbations on the semimajor axis $a$, on the other hand they change all the other Keplerian orbital elements with long-term harmonic shifts depending on the period of the perijove $\omega$ ($P_{\omega}\approx 500$ yr). 
In principle, the tidal effects of Saturn on the wide orbit of Juno might be of some relevance. An order-of-magnitude  evaluation of their importance can be performed by looking at the magnitude of the Kronian tidal acceleration on Juno. At the distance of Saturn from Jupiter expected at the beginning of the Juno science phase ($d_{\rm JS}=9.73$ au), the nominal Kronian tidal acceleration on Juno should be of the order of $A_{\rm tid}\approx Gm_{\rm S}\overline{r} d_{\rm JS}^{-3}= 2\times 10^{-11}$ m s$^{-2}$. It would yield a nominal displacement of about $\approx 10^4$ m over $T$. Thus, since the Saturn's gravitational parameter $Gm_{\rm S}$ is nowadays known with a fractional accuracy of the order of $\approx 3\times 10^{-8}$ \cite{2006AJ....132.2520J}, it can be reasonably concluded that the impact of the Kronian tides on the Juno's path is quite negligible for our purposes. Among the non-gravitational perturbations, whose impact on the proposed measurement is beyond the scope of this paper, particular attention should be paid to the drag from the Jovian atmosphere and to possible spurious thrusts from thermal out-gassing from the spacecraft potentially capable of affecting $a$.}
\subsection{The Earth's scenario}\lb{terra}
Moving to the Earth's gravitational field, it may be worth of mentioning that the  $\J2 c^{-2}$ effects cannot explain the flyby anomaly\footnote{\textcolor{black}{It occurred for artificial probes approaching the Earth along open, unbounded trajectories.}} \cite{2008PhRvL.100i1102A}. We performed a numerical analysis for the NEAR probe since it was most evident during the flyby of such a spacecraft occurred on 23 January 1998, when the X-Band Doppler apparatus detected an unexplained range-rate shift as large as $\Delta\dot\rho = 13.46\pm 0.01 $ mm s$^{-1}$ \cite{2008PhRvL.100i1102A}. We numerically integrated the equations of motion of NEAR with and without the $\J2 c^{-2}$ acceleration starting from the initial conditions of the 1998 flyby. The resulting $\J2 c^{-2}$ rang-rate
shift was as little as $\approx 10^{-6}$ mm s$^{-1}$.

A dedicated mission would have more chances, at least in principle. As an example, we considered a spacecraft endowed with a Doppler tracking apparatus moving along a highly elliptical ($e=0.79053$) polar orbit characterized by an altitude at the perigee of $h_{\rm min}=300$ km and a maximum geocentric distance as large as $r_{\rm max}=57103.2$ km. By numerically integrating its equations of motion in a geocentric equatorial ($\kx=\ky=0,\kz=1$) frame, it turned out that the magnitude of its $\J2 c^{-2}$ range-rate shift, referred to the geocenter, increases at every perigee pass in such a way that at, say, the 25th perigee passage  it amounts to $-0.4$ mm s$^{-1}$.
\section{Summary and conclusions}\lb{fine}
We re-examined the 1PN effects on the motion of a test particle orbiting an oblate central body in order to check if recent advancements in space science and technology  make them potentially measurable.

The answer is cautiously positive thanks to the ongoing Juno mission to Jupiter. Indeed, its peculiar orbital configuration, the accuracy of its radio-tracking apparatus and the huge Jovian oblateness make the perspective of measuring the $\J2 c^{-2}$ effects on the Juno's orbit feasible during its scheduled yearlong science phase in 2016-2017. Indeed, the numerically integrated $\J2 c^{-2}$ range-rate shifts at the perijove passes are expected to be as large as $\approx 280$ $\mu$m s$^{-1}$; the onboard Ka-Band Doppler system should be accurate at a $1-5$ $\mu$m s$^{-1}$ level during such passes. The signatures of other competing effects such as the classical shifts due to the even zonal harmonics of the Newtonian component of the aspherical gravitational potential of Jupiter are different, thus likely allowing for an adequate separation of the signal we are interested in. Most of the positive results of independent investigations previously made by other researchers in view of a possible measurement of the smaller Lense-Thirring effect with Juno  may be valid  for the $\J2 c^{-2}$ effects as well, thus  making us reasonably confident about a successful determination of them. Nonetheless,  a dedicated covariance analysis implying simulations of Doppler measurements and parameter estimation is required to further support the promising results of our preliminary sensitivity study.

We also looked at the Earth's gravitational field. Our numerical analysis for the NEAR spacecraft discards the possibility that the flyby anomaly can be due to the $\J2c^{-2}$ range-rate shift. We also investigated a hypothetical scenario based on the use of a dedicated spacecraft orbiting along a highly eccentric polar trajectory by finding that, for a perigee height of 300 km and an apogee geocentric distance of about 57000 km, the $\J2 c^{-2}$ range-rate shift would be of the order of $0.4$ mm s$^{-1}$. Such a figure should be confronted with the typical Doppler accuracy of about $0.01$ mm s$^{-1}$ reached in various past flybys by several spacecrafts.

\centering\bibliography{1PNJ2bib,PFEbib,satellitebib,historicbib,even_zonals_bib}{}
\end{document}